\RequirePackage{fix-cm}
\documentclass[smallextended]{svjour3}       
\smartqed  
\usepackage{amsmath, amsfonts, amssymb,bm,enumerate,multirow}
\usepackage{graphicx,psfrag,epsf,array, adjustbox,color,lineno}
\usepackage[english]{babel}

\newcommand{\half}{\frac{1}{2}}
\newcommand{\sigsq}{\sigma^2}
\newcommand{\sigsqhat}{\hat{\sigma}^2}
\newcommand{\bY}{\bm{Y}}

\newcommand{\ident}{\bm{I}}

\newcommand{\bX}{\bm{X}}
\newcommand{\bx}{\bm{x}}
\newcommand{\bZ}{\bm{Z}}
\newcommand{\bz}{\bm{z}}
\newcommand{\bu}{\bm{u}}

\newcommand{\bW}{\bm{W}}
\newcommand{\bepsilon}{\bm{\epsilon}}
\newcommand{\bmu}{\bm{\mu}}
\newcommand{\bbeta}{\bm{\beta}}

\newcommand{\Ytilde}{\widetilde{Y}}
\newcommand{\Xtilde}{\widetilde{X}}
\newcommand{\inv}{^{-1}}

\newcommand{\bV}{\bm{V}}
\newcommand{\bP}{\bm{P}}

\newcommand{\bU}{\bm{U}}
\newcommand{\bD}{\bm{D}}
\newcommand{\btheta}{\bm{\theta}}
\newcommand{\superth}{^{th}}

\journalname{}
\begin{document}

\title{An Approximate Restricted Likelihood Ratio Test for Variance Components in Generalized Linear Mixed Models 
}

\titlerunning{Approximate RLRT for Variance Components in GLMMs}        

\author{Stephanie T. Chen         \and
        Luo Xiao \and Ana-Maria Staicu 
}


\institute{Stephanie T Chen \at
              Department of Statistics, North Carolina State University \\
              \email{stchen3@ncsu.edu}
           \and Luo Xiao \at Department of Statistics, North Carolina State University \\ \email{lxiao5@ncsu.edu}
           \and Ana-Maria Staicu \at Department of Statistics, North Carolina State University \\
          \email{astaicu@ncsu.edu}
}

\date{Received: date / Accepted: date}

\maketitle

\begin{abstract}
		Generalized linear mixed models (GLMMs) are used to model responses from exponential families with a combination of fixed and random effects. For variance components in GLMMs, we propose an approximate restricted likelihood ratio test that conducts testing on the working responses used in penalized quasi-likelihood estimation. This presents the hypothesis test in terms of normalized responses, allowing for application of existing testing methods for linear mixed models. Our test is flexible, computationally efficient, and outperforms several competitors. We illustrate the utility of the proposed method with an extensive simulation study and two data applications. An \texttt{R} package is provided.
\end{abstract}

	\section{Introduction}
	Generalized linear mixed models (GLMMs) are widely used to model repeated observations from exponential family distributions. GLMMs build off generalized linear models (GLM) and linear mixed models (LMM) to model generalized responses with fixed and random effects (see \cite{McCullochetal2008} and \cite{Stroup2013} for an overview). As a result, GLMMs are able to accommodate scientifically relevant factors, such as subject-specific differences, and complex correlation structures. However, model estimation and inference becomes computationally more challenging with the presence of random effects. Thus, there is both scientific and computational interest in determining if random effects are truly necessary. In this paper, we consider testing random effects, or equivalently zero-value variance components, in GLMMs. 
	\\
	\\ For example, consider the iconic salamander mating experiment from \cite{McCullaghNelder1989}. Researchers are interested in factors that affect mating behavior, and may model binary mating success using a GLMM with population-level fixed effects and subject-specific random factors. Determining the significance of individual differences is as simple as testing the random effects in this GLMM.
	Another application is testing unspecified smooth functions in semiparametric or functional data models, such as in \cite{ZhangLin2003} and \cite{CrainiceanuRuppert2004}. Using the mixed effects representation of penalized splines to approximate smooth functions, a test of the functional form can be formulated in terms of random effects. Thus, there is need for accurate and flexible testing methods for random effects in GLMMs.
	\\
	\\ Testing variance components is difficult for GLMMs due to the lack of a closed-form likelihood for non-normal responses \cite{McCullochetal2008}. As a result, existing testing methods are closely tied to the availability of parameter estimation techniques. We briefly review several key estimation methods and related hypothesis tests. Penalized quasi-likelihood (PQL; \cite{Schall1991}, \cite{BreslowClayton1993}, \cite{WolfingerOconnel1993}) extends the quasi-likelihood approach for estimating GLMs \cite{McCullaghNelder1989}. 
	When applied to GLMMs, PQL allows for efficient parameter estimation but can return inaccurate estimates when the quasi-likelihood approximation is poor, and is also asymptotically biased estimates when applied to Bernoulli and Binomial responses with small denominators \cite{BreslowLin1995}, \cite{Bolkeretal2007}. A number of bias corrections have been proposed, but generally lack software implementations \cite{BreslowLin1995}, \cite{LeeNelder2001}. Because PQL maximizes quasi-likelihood rather than likelihood, it cannot be used with standard likelihood-based tests. Instead, \cite{Lin1997} proposes score tests for variance components in GLMMs by directly testing the ``normalized" working responses used for parameter estimation. This allows for calculation of the test statistic and its asymptotic null distribution. \cite{ZhangLin2003} extend these tests to generalized semiparametric additive models. However, these score tests have similar limitations to PQL, particularly for Bernoulli responses, for which the tests are conservative \cite{Lin1997}. Additionally, there is no standard software implementation, making application of the test infeasible for many practitioners.
	\\
	\\ Besides PQL, methods that directly approximate the likelihood (Laplace approximation \cite{Raudenbushetal2000}; Gauss-Hermite quadrature \cite{PinheiroChao2006}; Monte Carlo-based sampling \cite{McCulloch1997}, \cite{Knudson2016}), have been developed to allow for testing using standard methods. While these estimation methods are generally more accurate than PQL, they are also slower, less flexible, and cannot be applied to all models.
	When the likelihood can be directly approximated, \cite{MolenberghsVerbeke2007} compare likelihood ratio, score, and Wald tests for testing variance components, ultimately recommending likelihood ratio tests (LRTs) for their ease of implementation. However, because the null parameter value lies on the boundary of the parameter space, these tests rely on the nonstandard asymptotic null distribution \cite{SelfLiang1987}, a mixture of chi-square distributions. The asymptotic LRT is known to be conservative when applied to normal responses when (a) the sample size is small to moderate or (b) the assumption that responses can be divided into independent and identically distributed (iid) subvectors is violated \cite{PinheiroBates2000}, \cite{CrainiceanuRuppert2004}. \cite{ZhangLin2008} show that the LRT is conservative for testing binary responses, and we will show via simulation study that it is similarly conservative for Poisson and Binomial responses. While \cite{CrainiceanuRuppert2004} derive the finite sample null distribution for LMMs (normal responses) and show that it improves performance \cite{Grevenetal2008}, \cite{Scheipletal2008}, it remains unclear how to extend these results to generalized responses.
	Wald tests are also available and commonly used for testing fixed effects in GLMMs, but are highly conservative when applied to variance components and are strongly discouraged \cite{Stroup2013}. In summary, while asymptotic methods are available for testing variance components in GLMMs, these methods tend to be inaccurate and may be inapplicable for many scenarios.
	\\
	\\ To address these limitations, we propose an approximate restricted likelihood ratio test applied to the ``normalized" PQL working responses to test variance components in GLMMs. We calculate the test statistic from an approximate working LMM and compare it to the finite-sample null distribution from \cite{CrainiceanuRuppert2004}. This approach improves on the score tests developed by \cite{Lin1997} and \cite{ZhangLin2003}, and a user-friendly \texttt{R} implementation is provided.
	\\
	\\ The remainder of this paper is organized as follows. Section 2 presents the model and testing framework, and Section 3 describes the proposed methodology. Section 4 describes the software implementation. Section 5 presents a simulation study with comparison to three competing methods. Section 6 describes two data applications and Section 7 summarizes the paper.

	\section{Statistical Framework}
	Let $\bY$ be a vector of outcomes from an exponential family distribution with a known link function $g(x)$ and linear predictor $\bm{\eta}$, and assume that it follows the generalized linear mixed model (GLMM)
	\begin{equation}
		\begin{split}
			E(\bY | \bu_1, \dots, \bu_S) &= \bmu = g^{-1}(\bm{\eta})
			\\  \bm{\eta} & = \bX \bbeta + \sum_{s=1}^{S} \bZ_s\bu_s
			\\ \bu_s &\sim N(0,\sigsq_s\bD_s) \; \forall s= 1,\dots, S,
		\end{split}
	\end{equation}
	where $\bX$ is the design matrix for the fixed effects, $\bbeta$, and $\bZ_s$ is the design matrix for the $s\superth$ random effects vector $\bu_s$. We assume that each $q_s$-length random effects vector, $\bu_s$ has a shared variance component, $\sigsq_s$, and a known positive semi-definite matrix, $\bD_s$, and are independent between $s$. Thus, the individual outcomes of $\bY$ are independent conditional on the random effects.
	\\
	\\ We are interested in testing for the presence of a single random effect vector, $\bu_S$, or equivalently
	\begin{equation}
		H_0 \text{: } \sigsq_S = 0 \text{ vs } H_A \text{: } \sigsq_S > 0.
	\end{equation}
	\subsection{Examples}
	To illustrate the importance of the testing problem in (2), we describe three general cases. For simplicity, we only consider examples with one random effect, but variants or combinations of these basic structures are common in application (see \cite{McCullochetal2008}, and \cite{Stroup2013}, and \cite{Zuuretal2009} for examples). \\
	\\Example 1: Consider the $i\superth$ subject's $j\superth$ observation, $Y_{ij}$, modeled with a generalized mixed effects model and linear predictor $\eta_{ij}$ of the form
	\begin{align*}
		\eta_{ij} &= \bx_{ij}^T\bbeta + u_{i},
	\end{align*}
	where $\bx_{ij}$ is a vector of covariates for fixed effects $\bbeta$ and $u_{i} \sim N(0,\sigsq)$ are independent and identically distributed (iid) subject-specific random effects. In matrix form, $\bu  = (u_{1}, \dots, u_{n})^T$ is the vector of unique random subject intercepts with covariance $\sigsq\ident$, where $\ident$ is an identity matrix and $\bZ$ is the design matrix, which for the $i\superth$ column, has value 1 for observations from the $i\superth$ subject and 0 otherwise.
	\\
	\\ This model has been applied to the salamander mating study from \cite{McCullaghNelder1989}, where $Y_{ij}$ is the binary outcome for mating success between the $i\superth$ female and $j\superth$ male salamander, $\bx_{ij}$ are population-level characteristics, and $u_{i}$ is the random effect for the $i\superth$ female. For now, ignore the impact of individual male behavior. Modeling the probability of mating success using the above linear predictor accounts for individual behavior of the $i\superth$ female through the random effect, $u_i$. Determining if there is a significant difference in mating success between individual females is equivalent to testing if $\sigsq=0$.
	\\
	\\Example 2: Consider the $i\superth$ group's $j\superth$ observation, modeled with the generalized ANOVA-type model
	\begin{align*}
		\eta_{ij} &= \bx_{ij}^T\bbeta+ u_{i},
	\end{align*}
	where $\bx_{ij}$ is a vector of covariates for fixed effects $\bbeta$ and $u_{i} \sim N(0,\sigsq)$ are iid random group effects for a fixed number of groups, $n$. Note that in the previous example, the number of random effects, $n$, could increase with additional sampling. In matrix form, $\bu = (u_{1}, \dots, u_{n})^T$ is the vector of unique group intercepts with covariance $\sigsq\ident$ and design matrix $\bZ$, which, for the $i\superth$ column, has value 1 for observations from the $i\superth$ group and 0 otherwise. 
	\\
	\\This model has been used to study benthic species richness in the Netherlands (described in \cite{Zuuretal2009}), where $Y_{ij}$ is the Poisson-distributed species richness at the $j\superth$ coastal station in one of $i=1, \dots, 9$ intertidal areas (``beaches"), and $x_{ij}$ measures the amount of available food at that station. To determine if there are differences in species richness amongst these nine beaches, represented by the random beach effect, $u_i$, we can test if $\sigsq=0$. In this scenario, sampling additional stations does not increase the number of beaches (random effects).
	\\
	\\Example 3: Consider the nonparametric regression model from \cite{CrainiceanuRuppert2004}, where $Y_{ij}$ is the generalized response as previously described, and $\eta_{ij}$ is assumed to vary smoothly with covariate $t_{ij}$ such that $\eta_{ij} = f(t_{ij})$ for an unknown smooth function $f(t)$. Our goal is to test if $f(t)$ has a specific polynomial form, such as linear or quadratic. Equation (1) arises as a choice of modeling the unknown $f(t)$ as a combination of known basis functions with penalized coefficients. Specifically, we take the approach of \cite{Scheipletal2008} to model $f(t)$ using penalized spline bases with the mixed model
	\begin{align*}
	\eta_{ij} &= \bx_{ij}^T\bbeta + \bz_{ij}^T\bu,
	\end{align*}	
	where $\bbeta$ contains coefficients for polynomial basis functions and $\bu$ contains coefficients for non-polynomial basis functions. Thus, testing the form of $f(t)$ is equivalent to testing if the coefficients of the non-polynomial basis functions have zero variance. In this scenario, the number of random effects (dimension of $\bu$) depends only on the number of basis functions, and not directly on $n$ or $m$.
	\\
	\\ This model is used by \cite{ZhangLin2003} in a longitudinal study of childhood respiratory infections, where $Y_{ij}$ is a binary outcome for presence of infection for the $i\superth$ child at their $j\superth$ visit, $t_{ij}$ is the child's age, and $f(t)$ is a smooth function for the effect of age on risk for infection. The data suggests a strongly nonlinear effect of age. We can test this observation using the described mixed model framework and testing if $\sigsq=0$ for the nonlinear coefficients, $\bu$.
	
	\subsection{Likelihood of GLMMs}
	Standard methods for likelihood-based estimation and hypothesis testing are difficult to directly apply due to the lack of a closed-form GLMM likelihood. Specifically, let $\bu = (\bu_1,\dots,\bu_S)^T$ be the $\sum_{s}^S q_s$-dimensional vector of all random effects in equation (1), and denote by $f_{\bU}(\bu)$ its probability density function (pdf). If $f_{\bY|\bu}$ is the conditional pdf of $\bY$ corresponding to the exponential family model assumed for $\bY$, then the likelihood for equation (1) can be expressed as
	\begin{align*}
		\int f_{\bY|\bu}(\bm{y}|\bu) f_{\bU}(\bu) d\bu,
	\end{align*}
	which involves a $\sum_{s=1}^S q_s$-dimensional integral over the random effects. Numerical calculation is generally impractical, and the likelihood lacks a closed-form expression unless $\bY|\bu$ follows a normal distribution.
	\\
	\\ Methods such as Laplace approximation \cite{Raudenbushetal2000} and Gauss-Hermite quadrature {\cite{PinheiroChao2006}} can be used to approximate the likelihood, allowing for the use of standard testing methods. However, these estimation methods can be slow and inflexible, and their standard \texttt{R} implementation via the \texttt{glmer} function in the \texttt{lme4} package \cite{lme4} cannot be used for the nonparametric regression example described in Section 2.1. MCMC based methods are also popular due to their flexibility and have several \texttt{R} implementations (\texttt{MCMCglmm} \cite{MCMCglmm}, \texttt{glmm} \cite{glmm}). These methods are generally slow and difficult to generalize, and will not be considered in this paper. We will focus on penalized quasi-likelihood (PQL, \cite{Schall1991}, \cite{BreslowClayton1993}, \cite{WolfingerOconnel1993}) estimation because of its flexibility, computational efficiency, and convenient implementation in the \texttt{glmmPQL} function \cite{MASS}. However, PQL does not estimate the likelihood directly and cannot be used with standard hypotheses tests. In the following section, we propose a testing method that circumvents this limitation by conducting testing directly on the ``normalized" PQL working responses.
	
	\section{Methodology}
	\subsection{Overview}
	To test variance components in GLMMs, we propose an approximate restricted likelihood ratio test (RLRT) using the ``normalized" responses from PQL estimation. Briefly, the PQL algorithm iterates between (a) calculation of a ``normalized" working responses using parameter estimates to transform the generalized responses and (b) updating parameter estimates using a working LMM for the ``normalized" responses (see \cite{Schall1991}, \cite{BreslowClayton1993}, and \cite{WolfingerOconnel1993} for details). Our proposed test extends this approach by conducting testing on the PQL working responses at convergence, and consists of three steps: (a) transform the generalized responses to follow a normal distribution, (b) estimate the induced working LMM, and (c) use a RLRT to test the null hypothesis in equation (2). 
	\subsection{Proposed Test}		
	Following the PQL method for GLMM parameter estimation, consider the vector of standardized working responses
	\begin{equation}
		\bm{\Ytilde}= \bW^{*\half}[\bm{\eta}^*+ g'(\bmu^*)(\bY-\bmu^*)],
	\end{equation}
	where, at convergence, $\bm{\eta}^*$ is the linear predictor from equation (1), $g'(\bm{\mu}^*)$ is the derivative of the link function evaluated at the conditional mean, $\bmu^*$, and $\bW^* = \big[ \bU^T \bV^* \bU \big]\inv$ is a diagonal weight matrix where $\bU = \text{diag}\{g'(\bm{\eta}^*) \}$ and $\bV^* = Var(\bY|\bu_1,\dots,\bu_S)$ is the estimated diagonal conditional variance matrix. Note that $\bm{\Ytilde}$ is the standardized version of the typical working variate used in PQL estimation. We can estimate $\bm{\eta}^*$, $\bmu^*$, and $\bW^*$ using PQL estimates at convergence under the alternative hypothesis (see Section 4).
	\\
	\\ If $\bm{\eta}^*$, $\bmu^*$, and $\bW^*$ at convergence are assumed to be fixed, then $\bm{\Ytilde}$ is a linear function of $\bm{Y}$ and the likelihoods of the original and ``normalized" responses are proportional up to a constant. That is, $\bm{\Ytilde} = h(\bY)$, where $h(\cdot)$ is a linear function with slope $\bm{W}^{*\half}g'(\bmu^*)$ and intercept $\bm{W}^{*\half} [\bm{\eta}^* - g'(\bmu^*)\bmu^*]$, and the Jacobian of the transformation is the slope. The ``normalized" working responses can then be modeled with the working LMM
	\begin{equation}
		\begin{split}
			\bm{\Ytilde} &\approx \widetilde{\bX}\bbeta+\sum_{s=1}^{S} \widetilde{\bZ}_{s}\bu_{s} +\bepsilon
		\end{split}
	\end{equation}
	where $\widetilde{\bX}$ and $\widetilde{\bZ}_s$ are $\bX$ and $\bZ_s$ in equation (1) right-multiplied by $\bW^{*\half}$, respectively, $\bbeta$ and $\bu_{s}$ are as defined in equation (1), and $\bepsilon \sim N(0,\sigsq_e\ident)$, where $\ident$ is the identity matrix.
	\\
	\\ As a result, inference for the variance component $\sigsq_S$ in equation (4) is approximately equivalent to inference on $\sigsq_S$ in the original GLMM in equation (1). Since variance component estimates using restricted likelihood are generally better than those estimated using maximum likelihood \cite{PinheiroBates2000} and lead to better hypothesis testing performance \cite{Scheipletal2008}, we focus on restricted likelihood ratio tests. For testing the null hypothesis in (2), we propose an approximate restricted likelihood ratio test (\textit{aRLRT}) for the original generalized responses, $\bY$, that is equivalent to a RLRT for the ``normalized" responses ($\textit{aRLRT}_{\bm{Y}} = \textit{RLRT}_{\bm{\Ytilde}}$), of the form
	\begin{equation}
		\textit{aRLRT}_{\bY} = -2 \Big\{\sup_{\btheta \in H_0} \widetilde{\text{REL}}(\btheta) - \sup_{\btheta \in H_A} \widetilde{\text{REL}}(\btheta) \Big\},
	\end{equation}
	where $\widetilde{\text{REL}}(\btheta)$ denotes the restricted log-likelihood of the working LMM for $\bm{\Ytilde}$ and $\btheta = (\bbeta, \sigsq_1,\dots,\sigsq_S)^T$. In equation (5), $\widetilde{\text{REL}}(\btheta) = 
	-\half \Big[\log|\widetilde{\bV}| + \log| \widetilde{\bX}^T\widetilde{\bV}\widetilde{\bX}| + (N-p)\log(\bm{\Ytilde}^T\widetilde{\bP}^T\widetilde{\bV}\inv\widetilde{\bP}\bm{\Ytilde}) \Big]$, where $\widetilde{\bV} = Var(\bm{\Ytilde})$ 
	is the marginal variance of $\bm{\Ytilde}$, 
	$\widetilde{\bP} = \ident - \bm{\Xtilde}^T(\bm{\Xtilde}^T\widetilde{\bV}\inv\bm{\Xtilde})\inv \bm{\Xtilde}\widetilde{\bV}\inv$ is a projection matrix, $N$ is the total sample size, and $p$ is the dimension of $\bbeta$. We compare this statistic to the finite-sample null distribution derived in \cite{CrainiceanuRuppert2004}. Using a finite-sample distribution improves performance over the asymptotic null distribution from \cite{SelfLiang1987} and \cite{StramLee1994} by relaxing the assumptions that (a) responses can be divided into iid subvectors and (b) the number of subvectors tends to $\infty$. Notably, the former assumption is violated for the ANOVA and nonparametric regression models discussed in Section 2.1. However, this finite-sample result cannot be easily applied to the GLMM in equation (1) due to the difficulty of deriving and calculating the marginal model for generalized responses.
	
	\section{Implementation}
	To calculate the working responses $\bm{\Ytilde}$ in equation (3), we use PQL to estimate $\bm{\eta}^*$, $\bmu^*$, and $\bW^*$ at convergence under the alternative hypothesis with the \texttt{glmmPQL} function in \texttt{R} package \texttt{MASS} \cite{MASS}. Initial estimates can be obtained by setting the random effects to zero and estimating the fixed effects from the resulting generalized linear model. We make minor modifications to the \texttt{glmmPQL} function to use restricted maximum likelihood estimation and produce necessary output to standardize $\bm{\Ytilde}$. While we use PQL for estimation due to its fast computational speed and convenient interface, any method that estimates fixed and random effects can be used. The Supplementary Materials includes testing results using estimates from Laplace approximation via the \texttt{glmer} function in package \texttt{lme4} \cite{lme4}.
	\\
	\\ Equation (4) is then estimated under the null and alternative hypotheses using the \texttt{lme} function in package \texttt{nlme} \cite{nlme} to calculate the test statistic. The finite-sample null distribution is available in the \texttt{exactRLRT} function in package \texttt{RLRsim} \cite{Scheipletal2008}, and we compare the test statistic to 10,000 values from the finite sample distribution. An \texttt{R} package \texttt{glmmVCtest} implementing our method is available online \cite{glmmVCtest}.

	\section{Simulation Study}
	We conduct a simulation study to evaluate the performance of the proposed \textit{aRLRT} method compared with three competing methods, described in Section 5.2. Generate outcomes $Y_{ij}$ for $i=1, \dots, n$ and $j=1, \dots, m_i$ as
	\begin{align*}
		Y_{ij} = g^{-1}(\eta_{ij}),
	\end{align*}
	where $g(x)$ is the canonical link function and $\eta_{ij}$ is the linear predictor, to be described in Section 5.1. We consider a factorial combination of four factors: (a) distribution of $Y_{ij}$ [Normal, Bernoulli, Poisson, and Binomial (denominator = 4)]; (b) number of subjects or groups, $n$; (c) number of observations per subject or group, $m_i$; and (d) generating model. Details for the last three factors are given in Section 5.1. For each setting, we generate 5000 datasets for type I error rate and 1000 datasets for power.	
	
	\subsection{Generating Models}
		
	\subsubsection{M1. Random coefficient model - intercept}
	Consider a random coefficients model for the $i\superth$ group or subject's $j\superth$ observation, with fixed intercept $\beta_0$, fixed slope $\beta_1$, and random intercept $u_{0,i}\stackrel{iid}{\sim} N(0,\sigsq_0)$, of the form
	\begin{align*}
	\eta_{ij} = \beta_0 + \beta_1 x_{ij} + u_{0,i}.
	\end{align*}
	Our goal is to test if the random intercept is present, or $H_0: \sigsq_0 = 0$ versus $H_A: \sigsq_0 > 0$. Let $\beta_0 = 0$, $\beta_1 = 1$, $x_{ij} \sim Uniform[0,1]$, and vary $\sigsq_0 \ge 0$. We consider two specific models: (a) ANOVA-type with $n= 5, 10, 30$ groups and $m_i=m= 20, 100$ observations per group, and (b) linear mixed effects-type with $n= 20, 100$ subjects and $m_i=m= 5, 10, 30$ observations per subject.
		
	\subsubsection{M2. Random coefficient model - slope $|$ intercept}
	Consider a random coefficients model for the $i\superth$ group or subject's $j\superth$ observation, with fixed intercept $\beta_0$, fixed slope $\beta_1$, random intercept $u_{0,i} \stackrel{iid}{\sim} N(0,\sigsq_0)$, and random slope $u_{1,i} \stackrel{iid}{\sim} N(0,\sigsq_1)$, of the form
	\begin{align*}
	\eta_{ij} = \beta_0 + \beta_1 x_{ij} + u_{0,i} + u_{1,i}x_{ij}.
	\end{align*}
	This setting extends model M1 by adding a nuisance random effect. Our goal is to test if the random slope is present, or $H_0: \sigsq_1 = 0$ versus $H_A: \sigsq_1 > 0$. We use the same parameter settings as for model M2, with $\sigsq_0 =1$ and vary $\sigsq_1 \ge 0$. We consider two specific models: (a) ANOVA-type with $n= 5, 10, 30$ groups and $m_i=m= 20, 100$ observations per group, and (b) linear mixed effects-type with $n= 20, 100$ subjects and $m_i=m= 5, 10, 30$ observations per subject.		
		
	\subsubsection{M3. Linearity for nonparametric regression}
	Consider the nonparametric regression model $\eta_{ij} = f(t_{ij})$ for an unknown smooth function, $f(t)$, evaluated at covariate $t_{ij}$ for the $i\superth$ subject's $j\superth$ observation. We are interested in testing if $f(t)$ is a linear function, such that $f(t) = a + bt$ for some $a$ and $b$, against a nonlinear alternative, that is, $H_0: f(t)$ linear versus $H_A: f(t)$ nonlinear. To do so, we can use a penalized spline basis to re-frame this test in terms of variance components in a GLMM. Following \cite{Scheipletal2008}, let $B_k$ be B-splines with basis coefficients $\delta_k$ and second-order order penalty matrix $\bm{P}$ such that
	\begin{align*}
		\eta_{ij} & \approx \sum_{k=1}^K \delta_k B_k(t_{ij}) = \bx_{ij}^T\bbeta + \bz_{ij}^T\bu,
	\end{align*}
	where $\bbeta=[\beta_0,\beta_1]^T$ are fixed coefficients corresponding to linear basis functions, $\bx_{ij}$, and $\bu \sim N(\bm{0},\sigsq_S\ident_{K-2})$ are random coefficients corresponding to nonlinear basis functions $\bz_{ij}$. Thus, testing if $f(t)$ is a linear function is equivalent to testing $H_0: \sigsq_S = 0$ versus $H_A: \sigsq_S > 0$. Let $t_{ij} \sim Uniform[0,2]$ and $\delta \ge 0$.
	\\
	\\ For the simulation study, we follow \cite{ZhangLin2003} and let $f(t) = 0.5 - t + (0.25\delta)te^{2-2t}$ for $t \sim Uniform[0,2]$, where $\delta > 0$ is a scalar coefficient controlling deviation from the null hypothesis. We consider $n= 20, 100$ subjects and $m_i=m= 5, 10, 30$ observations per subject and use $K=30$ cubic B-splines for the approximation.
	
	\subsubsection{M4. Linearity for nonparametric regression $|$ random intercept}
	Consider the nonparametric regression model $\eta_{ij} = f(t_{ij}) + u_{0,i}$ for unknown smooth function, $f(t)$, and random intercept $u_{0,i} \stackrel{iid}{\sim} N(0,\sigsq_0)$. We are again interested in testing if $f(t)$ is a linear function against a nonparametric alternative, that is $H_0: f(t)$ linear versus $H_A: f(t)$ nonlinear, and can approximate $f(t)$ using the GLMM
	\begin{align*}
		\eta_{ij} & \approx \sum_{k=1}^K \delta_k B_k(t_{ij}) + u_{0,i}  = \bx^T\bbeta + \bz^T \bu + u_{0,i},
	\end{align*}
	where the terms are as defined for model M3. Thus, testing if $f(t)$ is a linear function is equivalent to testing $H_0: \sigsq_S = 0$ versus $H_A: \sigsq_S > 0$. We use the same parameter settings as for model M3, with $\sigsq_0 = 1$.
	
	\subsection{Competing Methods}
	We consider three additional likelihood-based methods for the hypothesis test in equation (2). The first two methods use the same procedure of conducting testing on the ``normalized" working responses and LMM in equation (4). The third method conducts testing directly on the generalized responses and GLMM. While Wald tests are also available and frequently used to test fixed effects in GLMMs, their use is strongly discouraged for testing variance components \cite{Stroup2013} and are not considered in this paper.
	
	\subsubsection{Asymptotic-Approximate Restricted Likelihood Ratio Test (\textit{as-aRLRT})}
	We consider an asymptotic variant of the proposed $aRLRT$ method by comparing the statistic in (5) to the asymptotic null distribution from \cite{SelfLiang1987}. When responses can be divided into iid subvectors tending to infinity, the test statistic for a single variance component follows a mixture of chi-square distributions, specifically $0.5 \chi^2_0 : 0.5 \chi^2_1$, where $\chi^2_0$ is a point mass at value zero. This assumption is violated for the ANOVA-type variants of models M1 and M2 when the number of groups is fixed, and the nonparametric regression models M3 and M4. We refer to this method as \textit{as-aRLRT}.
	
	\subsubsection{Approximate Score Test (\textit{aScore})}
	\cite{Lin1997} develop asymptotic score tests for global and individual variance components, and \cite{ZhangLin2003} extend their methods to testing smooth functions in semi-parametric additive models. Their method conducts testing on the working responses and LMM as described in Section 3, but instead of the likelihood ratio statistic, uses a score-based statistic with an asymptotic null distribution. We consider the bias-corrected Score test described in \cite{ZhangLin2003}, referred to as the \textit{aScore} test, using 10,000 samples from the asymptotic weighted chi-squared distribution.
	
	\subsubsection{Asymptotic Likelihood Ratio Test (\textit{asLRT})}
	\cite{MolenberghsVerbeke2007} recommends the likelihood ratio test (LRT) with the asymptotic null distribution from \cite{SelfLiang1987} for testing variance components in GLMMs. Unlike the previously described methods, this test is applied directly to the GLMM in equation (1). We use the \texttt{glmer} function in \texttt{R} package \texttt{lme4} \cite{lme4} for estimation, which uses higher-order Laplace approximation to calculate the GLMM likelihood. However, this function cannot be used for the nonparametric regression models M3 or M4 due to the lack of a grouping variable. Additionally, as mentioned for the \textit{as-aRLRT} method, the assumptions for the asymptotic null distribution are violated for the ANOVA-type variants of M1 and M2 when the number of groups is fixed. We refer to this method as \textit{asLRT}. For Normal responses where the likelihood can be directly calculated, this method is equivalent to the \textit{as-aRLRT} method using maximum likelihood instead of restricted maximum likelihood.	
	
	\subsection{Results}
	For conciseness, only type I error results for Normal, Bernoulli, and Poisson responses and power results for Bernoulli responses are shown in the main text; all others are included in the Electronic Supplementary Materials. 
	\subsubsection{Type I error}
	To compare performance between methods, we first consider the type I error rates for testing Normal responses (Table 1) where the ``normalizing" approximation is not required. The \textit{aRLRT} method maintains error rates close to the nominal level for $\alpha=0.05$ for all settings. The \textit{aScore} method is slightly conservative for model the linear mixed effects type-model M2, but otherwise maintains error rates close to $\alpha=0.05$. For normal responses, the \textit{as-aRLRT} and \textit{asLRT} methods are equivalent except for the use of restricted maximum likelihood versus maximum likelihood, respectively. The \textit{as-aRLRT} method has type I error rates closer to $\alpha=0.05$, but both methods are typically conservative. Testing Binomial responses (Table 1 in the Supplementary Materials) results in similar type I error rates to those observed for Normal responses.
		
	\begin{table}
		\caption{Empirical type I error rates for testing Normal responses at the nominal $\alpha=0.05$ level based on 5000 datasets, by generating model. The bolded term indicates the random effect or smooth function being tested. Legend: $n$: number of subjects or groups, $m$: number of observations per subject or group.}
		\label{tab:1}
		\begin{center}
		\scalebox{1}{\begin{tabular}{ l | c | c | c | c | c | c}
				Model: $i=1,\dots,n$, $j=1,\dots,m$ & $n$ & $m$ & \textit{aRLRT} & \textit{as-aRLRT} & \textit{aScore} & \textit{asLRT} \\ \hline
				\multirow{12}{*}{\textbf{M1}: $\beta_0 + \beta_1 x_{ij} + \bm{u_{0,i}}$ } 
				& 5 & 20 &  0.052 &	0.034 & 0.050 & 0.019 \\
				& 5 & 100 & 0.047 & 0.032 & 0.049 & 0.017 \\
				& 10 & 20 & 0.047 &	0.034 & 0.046 & 0.021 \\
				& 10 & 100 & 0.045 & 0.035 & 0.045 & 0.020 \\
				& 30 & 20 & 0.047 &	0.039 & 0.048 & 0.029 \\
				& 30 & 100 & 0.045 & 0.038 & 0.044 & 0.027 \\ \cline{2-7}
				& 20 & 5 & 0.051 & 0.046 & 0.047 & 0.033 \\
				& 20 & 10 & 0.045 &	0.037 & 0.044 & 0.029 \\
				& 20 & 30 & 0.047 &	0.038 &0.047 &  0.030 \\
				& 100 & 5 & 0.053 &	0.049 & 0.051 & 0.039 \\
				& 100 & 10 & 0.048 & 0.045 & 0.047 & 0.039 \\
				& 100 & 30 & 0.048 & 0.045 & 0.048 & 0.038 \\	\hline											
				\multirow{12}{*}{\textbf{M2}: $\beta_0 + \beta_1 x_{ij} + u_{0,i} + \bm{u_{1,i}}x_{ij}$ } 
				& 5 & 20 & 0.068 & 0.043 & 0.050 & 0.032 \\
				& 5 & 100 & 0.047 & 0.030 & 0.043 & 0.018 \\
				& 10 & 20 & 0.054 &	0.044 & 0.048 & 0.031 \\
				& 10 & 100 & 0.048 & 0.038 & 0.048 & 0.024 \\
				& 30 & 20 & 0.056 &	0.045 & 0.047 & 0.035 \\
				& 30 & 100 & 0.049 & 0.042 & 0.049 & 0.035 \\ \cline{2-7}								
				& 20 & 5 & 0.053 & 0.046 & 0.039 & 0.035 \\ 
				& 20 & 10 & 0.049 &	0.041 & 0.041 & 0.033 \\
				& 20 & 30 & 0.053 &	0.043 & 0.049 & 0.032 \\
				& 100 & 5 &  0.056 & 0.050 & 0.036 & 0.044 \\
				& 100 & 10 &  0.046 & 0.042 & 0.041 & 0.036 \\
				& 100 & 30 & 0.051 & 0.047 & 0.046 & 0.038 \\	\hline				
				\multirow{6}{*}{\textbf{M3}: $\bm{f(t_{ij})}$} 
				& 20 & 5 & 0.053 & 0.030 & 0.054 & 0.030 \\
				& 20 & 10 & 0.050 &	0.025 & 0.047 & 0.024 \\
				& 20 & 30 & 0.050 &	0.022 & 0.049 & 0.020 \\	
				& 100 & 5 &	0.052 &	0.025 & 0.050 & 0.023 \\
				& 100 & 10 & 0.052 & 0.026 & 0.047 & 0.024 \\
				& 100 & 30 & 0.050 & 0.023 & 0.045 & 0.020 \\ \hline														
				\multirow{6}{*}{\textbf{M4}: $b_{0,i} + \bm{f(t_{ij})}$} 
				& 20 & 5 & 0.054 & 0.026 & 0.051 & 0.026 \\
				& 20 & 10 & 0.046 &	0.023 & 0.047 & 0.022 \\
				& 20 & 30 & 0.052 &	0.029 &  0.049 & 0.027 \\	
				& 100 & 5 &	0.048 &	0.021 & 0.042 & 0.020 \\
				& 100 & 10 & 0.053 & 0.027 & 0.051 & 0.026 \\
				& 100 & 30 & 0.054 & 0.030 & 0.053 & 0.028
			\end{tabular}}
			\end{center}
		\end{table}	
		
	For Bernoulli responses, PQL is known to produce parameter estimates that become more biased as magnitude of the variance components increases \cite{BreslowLin1995}. As a result, the \textit{aRLRT} and \textit{aScore} methods maintain type I error rates for models without nuisance random effects (M1 and M3), but are somewhat inflated or conservative, respectively, for models M2 and M4 (Table 2). In particular, the \textit{aRLRT} method is inflated and the \textit{aScore} method is highly conservative for the linear mixed effects-type model M2 when $m$ is small, but improves with sample size. The \textit{as-aRLRT} method is conservative for all models that violate assumptions for the asymptotic null distribution (ANOVA-type M1 and M2, nonparametric regression M3 and M4). The \textit{asLRT} method is conservative for all models and sample sizes.
	\begin{table}
		\caption{Empirical type I error rates for testing Bernoulli (binary) responses at the nominal $\alpha=0.05$ level based on 5000 datasets, by generating model. The bolded term indicates the random effect or smooth function being tested. Legend: $n$: number of subjects or groups, $m$: number of observations per subject or group.}		\label{tab:2}
		\begin{center}
		\scalebox{1}{\begin{tabular}{ l | c | c | c | c | c | c}
					Model: $i=1,\dots,n$, $j=1,\dots,m$ & $n$ & $m$ & \textit{aRLRT} & \textit{as-aRLRT} & \textit{aScore} & \textit{asLRT} \\ \hline
					\multirow{12}{*}{\textbf{M1}: $\beta_0 + \beta_1 x_{ij} + \bm{b_{0,i}}$ } 
					& 5 & 20 & 0.051 & 0.032 & 0.058 & 0.020 \\
					& 5 & 100 & 0.044 &	0.028 & 0.046 & 0.014  \\
					& 10 & 20 & 0.045 &	0.035 & 0.049 & 0.022 \\
					& 10 & 100 & 0.051 & 0.040 & 0.052 & 0.024 \\
					& 30 & 20 & 0.047 &	 0.039 & 0.048 & 0.028 \\
					& 30 & 100 & 0.048 & 0.040 & 0.044 & 0.032 \\ \cline{2-7}	
					& 20 & 5 & 0.056 & 0.047 & 0.058 & 0.031  \\
					& 20 & 10 & 0.051 &	0.041 & 0.054 & 0.028 \\
					& 20 & 30 & 0.049 &	0.039 & 0.051 & 0.029  \\
					& 100 & 5 & 0.057 &	0.053 & 0.048 & 0.033  \\	
					& 100 & 10 & 0.058 & 0.052 & 0.053 & 0.037 \\
					& 100 & 30 & 0.052 & 0.049 & 0.052 & 0.039 \\	\hline											
					\multirow{12}{*}{\textbf{M2}: $\beta_0 + \beta_1 x_{ij} + b_{0,i} + \bm{b_{1,i}}x_{ij}$ } 
					& 5 & 20 & 0.062 & 	0.038 & 0.030 & 0.032  \\
					& 5 & 100 & 0.063 &	0.039 & 0.047 & 0.029 \\
					& 10 & 20 & 0.066 & 0.047 & 0.031 & 0.036  \\
					& 10 & 100 & 0.061 & 0.041 & 0.045 & 0.026 \\
					& 30 & 20 & 0.073 &	 0.062 & 0.040 & 0.036 \\
					& 30 & 100 & 0.064 & 0.052 & 0.056 & 0.033 \\ \cline{2-7}					
					& 20 & 5 & 0.063 & 0.052 & 0.014 & 0.036 \\ 
					& 20 & 10 & 0.071 &	0.057 & 0.022 & 0.038 \\
					& 20 & 30 & 0.057 &	0.044 & 0.040 & 0.031 \\
					& 100 & 5 & 0.074 &	0.068 & 0.012 & 0.032 \\
					& 100 & 10 & 0.084 & 0.078 & 0.027 & 0.027 \\
					& 100 & 30 & 0.076 & 0.067 & 0.048 & 0.031 \\	\hline						
					\multirow{6}{*}{\textbf{M3}: $\bm{f(t_{ij})}$} 
					& 20 & 5 & 0.044 & 0.021 & 0.048 & n/a \\	
					& 20 & 10 & 0.049 &	0.027 & 0.050 & n/a \\
					& 20 & 30 & 0.054 &	0.027 & 0.052 & n/a \\
					& 100 & 5 & 0.047 &	0.024 &	0.047 & n/a \\
					& 100 & 10 & 0.054 & 0.029 & 0.054 & n/a \\
					& 100 & 30 & 0.053 & 0.027 & 0.053 & n/a \\ \hline														
					\multirow{6}{*}{\textbf{M4}: $b_{0,i} + \bm{f(t_{ij})}$} 
					& 20 & 5 & 0.067 & 0.038 & 0.066 & n/a \\	
					& 20 & 10 & 0.060 &	0.030 & 0.058 & n/a \\
					& 20 & 30 & 0.056 &	0.030 & 0.051 & n/a \\
					& 100 & 5 &	0.064 &	0.034 & 0.062 & n/a \\
					& 100 & 10 & 0.059 & 0.030 & 0.057 & n/a \\
					& 100 & 30 & 0.062 & 0.032 & 0.057 & n/a
				\end{tabular}}
				\end{center}
			\end{table}
			
	In comparison, type I error rates for all methods are closer to the $\alpha=0.05$ level for testing Poisson responses, with the exception of models M2 (linear mixed effects type) and M4 (Table 3). For these models, error rates improve as $m$ increases. Additionally, PQL estimates are known to be poor when the mean (Poisson parameter) is small \cite{Bolkeretal2007}. This is reflected by the improvement in error rates for testing model M2 when the mean ($\beta_0$) increases from 0 to 2. Again, the \textit{as-aRLRT} method is conservative when assumption for the asymptotic null distribution are violated (ANOVA-type M1 and M2, nonparametric regression M3 and M4), and the \textit{asLRT} method is conservative for all models and sample sizes.
	\begin{table}
		\caption{Empirical type I error rates for testing Poisson responses at the nominal $\alpha=0.05$ level based on 5000 datasets, by generating model. The bolded term indicates the random effect or smooth function being tested. Legend: $n$: number of subjects or groups, $m$: number of observations per subject or group.}
		\label{tab:3}
		\begin{center}
		\scalebox{1}{\begin{tabular}{ l | c | c | c | c | c | c}
				Model: $i = 1, \dots, n$, $j=1,\dots,m$ & $n$ & $m$ & \textit{aRLRT} & \textit{as-aRLRT} & \textit{aScore} & \textit{asLRT} \\ \hline
						\multirow{12}{*}{\textbf{M1}: $\beta_0 + \beta_1 x_{ij} + \bm{b_{0,i}}$ } 
						& 5 & 20 & 0.057 &	0.040 & 0.057 &  0.019 \\
						& 5 & 100 & 0.046 &	0.030 & 0.045 & 0.014 \\
						& 10 & 20 & 0.052 &	0.039 & 0.049 & 0.018 \\
						& 10 & 100 & 0.050 & 0.038 & 0.054 & 0.027 \\
						& 30 & 20 & 0.054 &	0.046 & 0.053 & 0.030 \\
						& 30 & 100 & 0.045 & 0.037 & 0.046 & 0.028 \\ \cline{2-7}
						& 20 & 5 & 0.058  &	0.053 & 0.057 & 0.023  \\
						& 20 & 10 & 0.060  &	0.050 & 0.061 &  0.028  \\
						& 20 & 30 & 0.053  &	0.043 & 0.052 & 0.028 \\
						& 100 & 5 & 0.072  &	0.069 & 0.072 & 0.041 \\
						& 100 & 10 & 0.061 &	0.056 & 0.060 & 0.041 \\
						& 100 & 30 & 0.054 &	0.047 & 0.054 & 0.041 \\ \hline													
						\multirow{12}{*}{\textbf{M2}: $0 + \beta_1 x_{ij} + b_{0,i} + \bm{b_{1,i}}x_{ij}$ } 
						& 5 & 20 & 0.061  &	0.040 & 0.044 & 0.020 \\
						& 5 & 100 & 0.054 &	0.031 & 0.050 & 0.015 \\
						& 10 & 20 & 0.058  &	0.041 & 0.051 & 0.020 \\
						& 10 & 100 & 0.055  &	0.036 & 0.050 & 0.018 \\
						& 30 & 20 & 0.062 &	0.049 & 0.055 & 0.027 \\
						& 30 & 100 & 0.052 &	0.042 & 0.052 & 0.029 \\ \cline{2-7}			
						& 20 & 5 &  0.083  &	0.068 &  0.058 & 0.030 \\
						& 20 & 10 &  0.078  &	0.063 & 0.059 & 0.026 \\
						& 20 & 30 & 0.058 &	0.043 & 0.054 & 0.022 \\
						& 100 & 5 &  0.123  &	0.105 & 0.093 & 0.035 \\
						& 100 & 10 &  0.094 &	0.080 & 0.082 & 0.032 \\
						& 100 & 30 &  0.065 &	0.053 & 0.064 & 0.034 \\	\hline
						\multirow{6}{*}{\textbf{M2}: $2 + \beta_1 x_{ij} + b_{0,i} + \bm{b_{1,i}}x_{ij}$} 
						& 20 & 5 & 0.057  &	0.045  & 0.047 & 0.023 \\
						& 20 & 10 &  0.059 & 0.045 & 0.054 & 0.026 \\
						& 20 & 30 & 0.050  & 0.037 & 0.052 & 0.024 \\
						& 100 & 5 & 0.060  & 0.051 & 0.052 & 0.028 \\
						& 100 & 10 &  0.057  & 0.047 & 0.052 & 0.035 \\
						& 100 & 30 & 0.059  & 0.048 & 0.057 & 0.032 \\ \hline																		
						\multirow{6}{*}{\textbf{M3}: $\bm{f(t_{ij})}$} 
						& 20 & 5 &  0.052  &	0.029 & 0.050 & n/a \\	
						& 20 & 10 & 0.049  &	0.026 & 0.049 & n/a \\
						& 20 & 30 & 0.050  &	0.029 & 0.047 &  n/a \\
						& 100 & 5 & 0.051  &	0.026 &	0.047 & n/a \\
						& 100 & 10 & 0.055  &	0.027 & 0.055 & n/a \\
						& 100 & 30 & 0.051  &	0.026 & 0.048 & n/a \\ \hline														
						\multirow{6}{*}{\textbf{M4}: $b_{0,i} + \bm{f(t_{ij})}$} 
						& 20 & 5 &  0.091  &	0.054 & 0.075 & n/a \\	
						& 20 & 10 & 0.077  &	0.040 & 0.071 & n/a \\
						& 20 & 30 & 0.059 &	0.033 & 0.055 & n/a \\
						& 100 & 5 &	0.090  &	0.050 & 0.076 & n/a \\
						& 100 & 10 & 0.071  & 0.039 & 0.067 & n/a \\
						& 100 & 30 & 0.061  & 0.033 & 0.056 & n/a
					\end{tabular}}
					\end{center}
		\end{table}
				
	\subsubsection{Power}
	While we only show results for Bernoulli responses (Figure 1) in the main text, we observe similar performance patterns for Normal, Binomial, and Poisson responses (Figures 1, 2, 3 in Supplementary Materials). For all methods, power is higher when testing models without nuisance random effects (models M1 and M3) and increases with sample size (both $n$ and $m$). The \textit{aRLRT} method has similar or higher power than all other methods. The \textit{aScore} test has comparable power to the \textit{aRLRT} method for testing most models, but can have 5-20\% lower power for the random coefficients model M2. The \textit{as-aRLRT} and \textit{asLRT} methods have consistently lower power. For models M3 and M4 when the sample size is small, power may not converge to 100\% as deviation from the null hypothesis increases. For example, power peaks at 95\% when testing model M3 with $n=100$ subjects and $m=5$ observations per subject. In these scenarios, the probability of Bernoulli events converges to 0\% and/or 100\%, making logistic regression and hypothesis testing unsuitable for the data. This issue occurs only when testing Bernoulli responses, and power is higher for all methods when applied to Normal, Binomial, or Poisson responses (Supplementary Materials).
	\begin{figure}
		\begin{center}
		\includegraphics[scale=0.11]{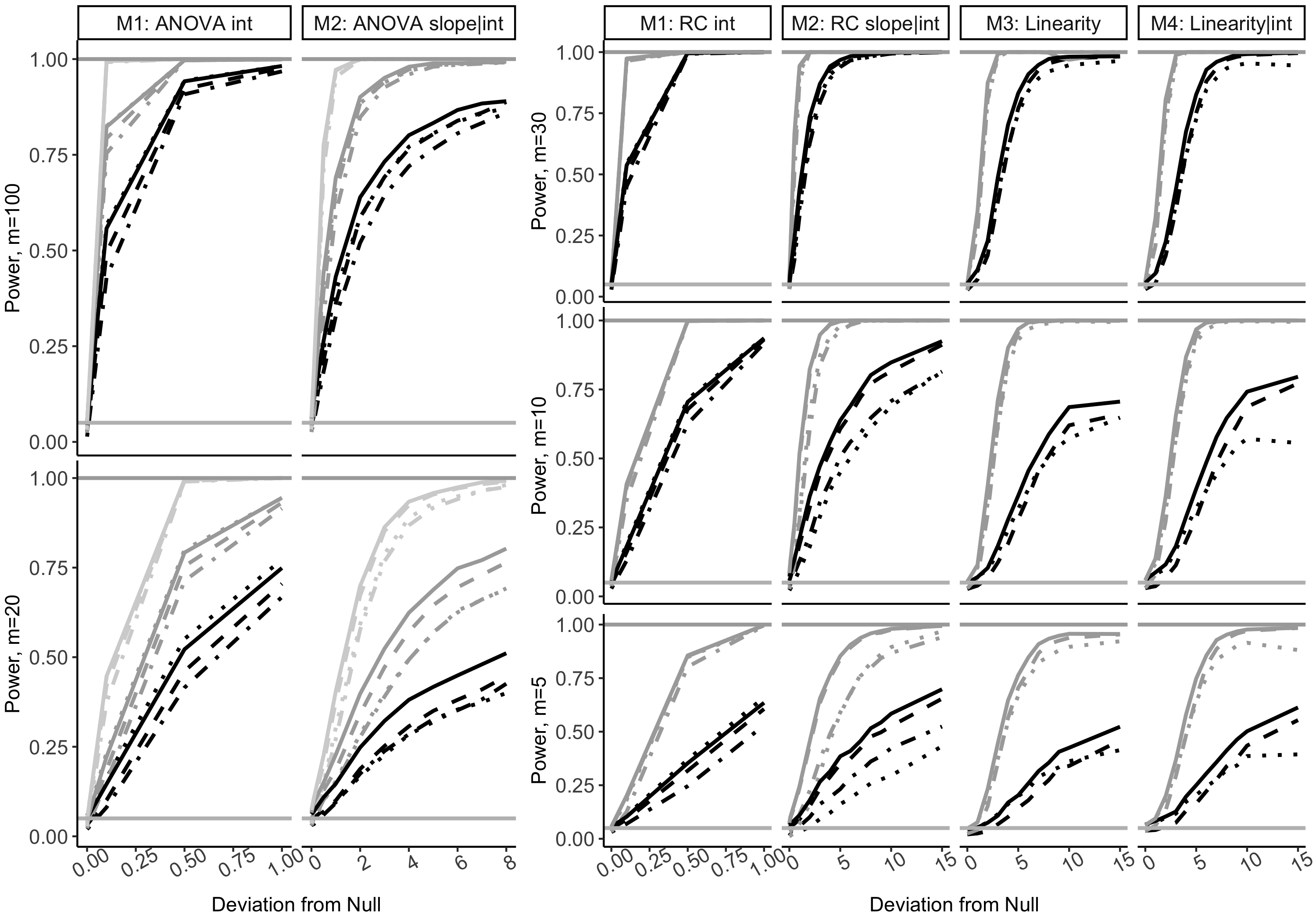}	
		\end{center}			
		\caption{Power for Bernoulli responses at the $\alpha=0.05$ level based on 1000 simulated datasets, by simulation model. Legend: \textit{aRLRT} (solid), \textit{aScore} (short dash), \textit{as-aRLRT} (long dash), \textit{asLRT} (short \& long dash). Right plots: $n=5$ groups (black), $n=10$ groups (dark gray), $n=30$ groups (light gray). Left plots: $n=20$ subjects (black), $n=100$ subjects (gray).}
		\label{fig:1}
	\end{figure}
				
	\subsection{Summary}
	Existing likelihood-based methods for testing variance components in GLMMs may have conservative type I error rates and low power for typical models. Conducting testing on the ``normalized" responses from PQL estimation allows for use of finite-sample null distributions to improve testing performance, particularly for small sample sizes. This approach outperforms existing methods applied directly to generalized responses. However, there were two scenarios where PQL estimation is known to be unreliable that impacts hypothesis testing performance. First, Bernoulli models with nuisance random effects may have biased estimates that lead to somewhat inflated type I error rates (6-8\% instead of 5\%). A related issue may occur for small sample sizes when the probability of Bernoulli events converges to 0\% or 100\%, so power does not reach 100\%. Second, estimates for Poisson models with a combination of (a) nuisance random effects, (b) small means (high proportion of zero-values), and (c) small sample size can be unreliable, leading to inflated type I error rates (6-11\%). Caution should be used for testing in these scenarios regardless of method used. Overall, the proposed \textit{aRLRT} method is fast, flexible, and has good performance for the range of models, distributions, and sample sizes considered in this study.

\section{Applications}
\subsection{Salamander Mating Behavior}
We return to the salamander mating study described in Sections 1 and 2.1, available in the \texttt{glmm} package \cite{glmm}. Briefly, 60 female and 60 male salamanders from Rough Butt (R) and White Side (W) populations were paired to determine if mating was more likely to occur between individuals from the same population. Individuals were paired for a total of 90 trials for each of the R/W and female/male combinations, for a total of 360 binary mating outcomes (see \cite{McCullaghNelder1989} for study details). We consider the GLMM from \cite{KarimZeger1992} and \cite{Knudson2016} for $Y_{ij}$, the binary mating outcome between the $i\superth$ female and $j\superth$ male salamander 
\begin{equation}
	\begin{split}
		Y_{ij} & = g\inv(\eta_{ij})
		\\ \eta_{ij} & = \bx_{ij}^T\bbeta + u_{1,i} + u_{2,j},
	\end{split}
\end{equation}
where $g(x)$ is the logit function, $\bx_{ij}$ is a vector of indicators for the population crosses $\bbeta = (\beta_{R/R}, \beta_{R/W}, \beta_{W/W}, \beta_{W/R})^T$, $u_{1,i} \stackrel{iid}{\sim} N(0,\sigsq_1)$ is the random subject-specific female effect, and $u_{2,j} \stackrel{iid}{\sim} N(0,\sigsq_2)$ is the random subject-specific male effect. We consider testing the significance of a male effect in the presence of a female effect, or $H_0: \sigsq_2 = 0$ versus $H_A: \sigsq_2 > 0$.
\\
\\ The parameter estimates for equation (6) using PQL are $\hat{\bbeta} = (0.930,  0.283,  0.903, -1.801)^T$, $\sigsqhat_1 = 1.201$, and $\sigsqhat_2 = 1.142$. The \textit{aRLRT} and \textit{as-aRLRT} methods estimate a test statistic of $17.074$, \textit{aScore} estimates a statistic of $81.080$, and the \textit{asLRT} estimates a statistic of $11.685$, all corresponding to $p<0.001$. Thus, all four methods reject the null hypothesis, indicating that individual male salamanders have different rates of mating success.

\subsection{Benthic Species Richness in the Netherlands}
We return to the species richness study described in Section 2.1 from \cite{Zuuretal2009}. Let $Y_{ij}$ be the species richness (assumed to follow a Poisson distribution) at the $j\superth$ location in the $i=1, \dots, 9$ beach. Our goal is to determine if there are significant differences in species richness across the nine beaches. Consider the GLMM
\begin{equation}
	\begin{split}
		Y_{ij} & = g\inv(\eta_{ij})
		\\ \eta_{ij} &= \beta_0 + \beta_1x_{ij} + u_{i},
	\end{split}
\end{equation}
where $g(x)$ is the log function, $\beta_0$ is a fixed intercept, $x_{ij}$ is the Normal Amsterdams Peil (NAP), a measure of available food with slope $\beta_1$, and $u_{i} \stackrel{iid}{\sim} N(0,\sigsq)$ is the random beach effect for the $i\superth$ beach. To test for a significant difference in richness across beaches, we test $H_0: \sigsq = 0$ versus $H_A: \sigsq > 0$.
\\
\\ The parameter estimates for equation (7) using PQL are $\beta_0 = 1.684$, $\beta_1 = -0.504$, and $\sigsq_1 = 0.492$. The \textit{aRLRT} and \textit{as-aRLRT} methods estimate a test statistic of $16.654$, the \textit{aScore} test estimates a statistic of $1015.228$, and the \textit{asLRT} method estimates a statistic of $40.396$, all corresponding to $p<0.001$. All four methods indicate that species richness differs significantly across the nine beaches.

\section{Concluding Remarks}
In this paper, we propose an approximate restricted likelihood ratio test for variance components in generalized linear mixed models and develop an \texttt{R} package \cite{glmmVCtest} for easy implementation. Our method extends the PQL framework to conduct testing on a ``normalized" working response and linear mixed model. This allows for testing using restricted likelihood and a finite-sample null distribution with existing results and software. We find that the proposed \textit{aRLRT} method is computationally efficient and has better performance than three competitor methods for testing Normal, Bernoulli, Poisson, and Binomial responses for a range of common models and settings. In particular, all approximate methods applied to ``normalized" responses outperformed asymptotic methods applied directly to the generalized responses. However, the method can have inflated type I error rates in two scenarios when PQL estimation is known to be poor: (a) models for Bernoulli responses with nuisance random effects and (b) Poisson data with a small mean and sample size.

\section*{Supplementary Materials}
The Electronic Supplementary Materials referenced in Sections 4 and 5 are available online with this article and contain additional simulation results for testing Normal, Binomial, and Poisson responses. It also includes alternative results using Laplace approximation to estimate $\bm{\Ytilde}$. An \texttt{glmmVCtest} package implementing all methods is available online \cite{glmmVCtest}.
	
\begin{acknowledgements}
The authors thank Dr. Daowen Zhang for providing code to conduct the \textit{aScore} test. ST Chen and L Xiao's research were supported by grant numbers OPP1148351
and OPP1114097 from the Bill and Melinda Gates Foundation. AM Staicu's research was
supported by National Science Foundation grant number DMS 1454942 and National Institute of Health grants 5P01 CA142538-09 and 2R01MH086633. This work represents the opinions of the researchers and not necessarily that of the granting organizations.
\end{acknowledgements}

\bibliographystyle{spmpsci}      
\bibliography{GLMMbib}   

\end{document}